\documentclass[journal=jctcce,manuscript=article]{achemso}

\renewcommand{\phi}{\varphi}
\renewcommand{\theta}{\vartheta}
\renewcommand{\iota}{\ensuremath{\imath}}
\renewcommand{\epsilon}{\varepsilon}

\newcommand{\distance}{\ensuremath{\mathnormal{r}}}
\newcommand{\distancecut}{\ensuremath{\distance_\mathrm{c}}}

\newcommand{\kboltz}{\ensuremath{\mathnormal{k}_\mathrm{B}}}

\usepackage[british]{babel}

\usepackage{epsfig}
\usepackage{amsmath}
\usepackage{amssymb}
\usepackage{subfig}
\usepackage{xspace}

\newcommand*{\bohr}{\ensuremath{\mathnormal{a}_\mathrm{H}}}
\newcommand*{\elementarycharge}{\ensuremath{\mathrm{e}}}

\newcommand*{\SI}[2]{\textnormal{{#1} {#2}}}
\newcommand*{\mardyn}{\textit{ls1 mardyn}\xspace}

\graphicspath{ {./images/} }

\author{Christoph Niethammer}
\affiliation[High Performance Computing Center Stuttgart (HLRS), Germany]{High Performance Computing Center Stuttgart, Nobelstr.\ 19, 70569 Stuttgart, Germany}
\author{Stefan Becker}
\affiliation[Laboratory of Engineering Thermodynamics (LTD), Univ.\ of Kaiserslautern, Germany]{University of Kaiserslautern, Laboratory of Engineering Thermodynamics, Erwin-Schr\"odinger-Str.\ 44, 67663 Kaiserslautern, Germany}
\author{Martin Bernreuther}
\affiliation[High Performance Computing Center Stuttgart (HLRS), Germany]{High Performance Computing Center Stuttgart, Nobelstr.\ 19, 70569 Stuttgart, Germany}
\author{Martin Buchholz}
\author{Wolfgang Eckhardt}
\author{Alexander Heinecke}
\affiliation[Scientific Computing in Computer Science (SCCS), TU M\"unchen, Germany]{TU M\"{u}nchen, Chair for Scientific Computing in Computer Science, Boltzmannstr.\ 3, 85748 Garching, Germany}
\author{Stephan Werth}
\affiliation[Laboratory of Engineering Thermodynamics (LTD), Univ.\ of Kaiserslautern, Germany]{University of Kaiserslautern, Laboratory of Engineering Thermodynamics, Erwin-Schr\"odinger-Str.\ 44, 67663 Kaiserslautern, Germany}
\author{Hans-Joachim Bungartz}
\affiliation[Scientific Computing in Computer Science (SCCS), TU M\"unchen, Germany]{TU M\"{u}nchen, Chair for Scientific Computing in Computer Science, Boltzmannstr.\ 3, 85748 Garching, Germany}
\author{Colin W. Glass}
\affiliation[High Performance Computing Center Stuttgart (HLRS), Germany]{High Performance Computing Center Stuttgart, Nobelstr.\ 19, 70569 Stuttgart, Germany}
\author{Hans Hasse}
\affiliation[Laboratory of Engineering Thermodynamics (LTD), Univ.\ of Kaiserslautern, Germany]{University of Kaiserslautern, Laboratory of Engineering Thermodynamics, Erwin-Schr\"odinger-Str.\ 44, 67663 Kaiserslautern, Germany}
\author{Jadran Vrabec}
\affiliation[Thermodynamics and Energy Technology (ThEt), Univ.\ of Paderborn, Germany]{University of Paderborn, Laboratory of Thermodynamics and Energy Technology, Warburger Str.\ 100, 33098 Paderborn, Germany}
\author{Martin Horsch}
\affiliation[Laboratory of Engineering Thermodynamics (LTD), Univ.\ of Kaiserslautern, Germany]{University of Kaiserslautern, Laboratory of Engineering Thermodynamics, Erwin-Schr\"odinger-Str.\ 44, 67663 Kaiserslautern, Germany}
\email{martin.horsch@mv.uni-kl.de}

\title{\mardyn: The massively parallel molecular dynamics code for large systems}

\begin{document}

\begin{abstract}
The molecular dynamics simulation code \mardyn is presented.
It is a highly scalable code, optimized for massively parallel execution on supercomputing architectures, and currently holds the world record for the largest molecular simulation with over four trillion particles.
It enables the application of pair potentials to length and time scales which were previously out of scope for molecular dynamics simulation.
With an efficient dynamic load balancing scheme, it delivers high scalability even for challenging heterogeneous configurations.
Presently, multi-center rigid potential models based on Lennard-Jones sites, point charges and higher-order polarities are supported.
Due to its modular design, \mardyn can be extended to new physical models, methods, and algorithms, allowing future users to tailor it to suit their respective needs.
Possible applications include scenarios with complex geometries, e.g.\ for fluids at interfaces, as well as non-equilibrium molecular dynamics simulation of heat and mass transfer.
\end{abstract}

\section{Introduction}
\label{sec:intro}

The molecular dynamics (MD) simulation code \mardyn (large systems 1: molecular dynamics) is presented here. The \mardyn program is an interdisciplinary endeavor, whose contributors have backgrounds from engineering, computer science and physics, aiming at studying challenging scenarios with up to trillions of molecules. In the considered systems, the spatial distribution of the molecules may be heterogeneous and subject to rapid unpredictable change. This is reflected by the algorithms and data structures as well as a highly modular software engineering approach. The source code of \mardyn is made publicly available as free software under a two-clause BSD license.\cite{Website}

Molecular modelling and simulation has become a powerful computational method\cite{AT87, Frenkel2002} and is applied to a wide variety of areas such as thermodynamic properties of fluids,\cite{Deublein2011} phase equilibria,\cite{Moeller1990, Vrabec2002} interfacial properties,\cite{Rusanov1977} phase transitions,\cite{Rao1978, Angelil2014} transport coefficients,\cite{Chialvo1991} adsorption, \cite{Sokolowski1990, Horsch2010} mechanical properties of solids,\cite{Roesch2009} flow phenomena,\cite{Thompson1997, Frentrup2012} polymer properties,\cite{MuellerPlathe2002} protein folding, \cite{Lee2009, Lindorff2011} or self-assembly.\cite{Engel2007}
The sound physical basis of the approach makes it extremely versatile.
For a given force field, the phase space can be explored by molecular dynamics simulation under a variety of boundary conditions, which allows gathering information on all thermodynamic states and processes on the molecular level. If required, external forces (e.g.\ an electric field) can be imposed in addition to the intermolecular interactions.

MD simulation has an extremely high temporal and spatial resolution of the order of $10^{-15}$ seconds and $10^{-11}$ meters, respectively. This resolution is useful for studying physical phenomena at small length scales, such as the structure of fluid interfaces. With a time discretization on the femtosecond scale, rapid processes are immediately accessible, while slower processes may require particularly devised sampling techniques such as metadynamics.\cite{Laio2002}
The number of molecules is also a challenge for molecular simulation.
While systems of practical interest contain extremely large numbers of molecules, e.g.\ of the order of $10^{23}$, the largest ensembles that can be handled today are of the order of $10^{12}$ molecules.\cite{Eckhardt-2013}
This limitation is usually addressed by focusing on representative subvolumes, containing a limited number of molecules, to which an appropriate set of boundary conditions is applied.
Depending on the type of information that is determined, e.g.\ transport properties \cite{Guevara2011} or phase equilibria \cite{Moeller1990,Vrabec2002} of bulk phases, a number of molecules of the order of $1$ $000$ may be sufficient.
However, non-equilibrium scenarios such as condensation \cite{Horsch2008, Horsch2009} or mass transfer through nanoporous membranes\cite{Frentrup2012, Mueller13} require much larger simulation volumes.


There are so many scalable MD codes available that a comprehensive discussion would be beyond the scope of the present paper. For the development of MD codes, as for any software, there are trade-offs between generality and optimization for a single purpose, which no particular implementation can completely evade.
Several popular MD simulation environments are tailored for studying biologically relevant systems, with typical application scenarios including conformational sampling of macromolecules in aqueous solution. The relaxation processes of such systems are often several orders of magnitude slower than for simple fluids, requiring an emphasis on sampling techniques and long simulation times, but not necessarily on large systems.

The \textit{AMBER} package,\cite{Case2005} for instance, scales well for systems containing up to $400$ $000$ molecules, facilitating MD simulations that reach the microsecond time scale.\cite{Salomon2013} Similarly, \textit{GRO\-MACS}\cite{Berendsen1995, Pronk2013} and \textit{NAMD},\cite{Phillips2005} which also have a focus on biosystems, have been shown to perform quite efficiently on modern HPC architectures as well. \textit{Tinker} was optimized for biosystems with polarizable force fields,\cite{Ren2011} whereas \textit{CHARMM},\cite{Brooks1983} which was co-developed by Nobel prize winner Martin Karplus, is suitable for coupling classical MD simulation of macromolecules with quantum mechanics.\cite{Brooks2009} 
The \textit{LAMMPS} program \cite{Plimpton1995c, Brown2011, Plimpton2012, Diemand2013} as well as \textit{DL\_POLY},\cite{Todorov2006} which scales well for homogeneous fluid systems with up to tens of millions of molecules, and \textit{ESPResSo},\cite{Limbach2006} which emphasizes its versatility and covers both molecular and mesoscopic simulation approaches, are highly performant codes which aim at a high degree of generality, including many classes of pair potentials and methods. The \textit{ms2} program performs well for the simulation of vapor-liquid equilibria and other thermodynamic properties,\cite{Deublein2011} but is limited to relatively small numbers of molecules. The \textit{IMD} code,\cite{Stadler1997, Roth2000} which has twice before held the MD simulation world record in terms of system size, has a focus on multi-body potentials for solids.


With \mardyn{}, which is presented here, a novel MD code is made available to the public. It is more specialized than most of the molecular simulation programs mentioned above. In particular, it is restricted to rigid molecules, and only constant volume ensembles are supported, so that the pressure cannot be specified in advance. Electrostatic long-range interactions, beyond the cut-off radius, are considered by the reaction field method,\cite{Saager1991} which cannot be applied to systems containing ions. However, \mardyn{} is highly performant and scalable. Holding the present world record in simulated system size,\cite{Eckhardt-2013} it is furthermore characterized by a \textit{modular structure}, facilitating a high degree of flexibility within a \textit{single code base}.
Thus, \mardyn is not only a simulation engine, but also a framework for developing and evaluating simulation algorithms, e.g.\ different thermostats or parallelization schemes. Therefore, its software structure supports alternative implementations for methods in most parts of the program, including core parts such as the numerical integration of the equations of motion. The C++ programming language was used, including low level optimizations for particular HPC systems. In this way, \mardyn has been proven to run efficiently on a variety of architectures, from ordinary workstations to massively-parallel supercomputers.

In a fluid system, neighborhood relations between molecules are always subject to rapid change.
Thus, the neighbor molecules have to be redetermined throughout the simulation.
For this purpose, \mardyn{} employs a linked-cell data structure,\cite{Quentrec1973, Hockney1981, Schamberger2003} which is efficiently parallelized by spatial domain decomposition.\cite{Bernreuther2005a, Bernreuther2008}
Thereby, the simulation volume is divided into subvolumes that are assigned to different processes.
Interactions with molecules in adjacent subvolumes are explicitly accounted for by synchronized halo regions.\cite{Frenkel2002}

Using \mardyn{}, a wide range of simulation scenarios can be addressed,
and pre-release versions of \mardyn{} have already been successfully applied to a variety of topics from chemical and process engineering: Nucleation in supersaturated vapors \cite{Horsch2008b, Horsch2009, Vrabec2009b, Horsch2011} was considered with a particular focus on systems with a large number of particles.\cite{Grottel2007, Horsch2008, Horsch2009b} On the SuperMUC, over four trillion molecules were simulated.\cite{Eckhardt-2013} The vapor-liquid surface tension and its dependence on size and curvature was characterized.\cite{HHSAEVMJ2012, Vrabec2009b, WLHH2013, HH14, Werth2014} The \mardyn program was furthermore employed to investigate fluid flow through nanoporous membrane materials \cite{HVBH2009} and adsorption phenomena such as the fluid-solid contact angle in dependence on the fluid-wall interaction.\cite{Horsch2010, HNVH13}

Scenario generators for \mardyn{} are available both internally, i.e.\ without hard disk input/output, and as external executables. The internal generators create the initial configuration directly in memory, which is distributed among the processes, facilitating a better scalability for massively-parallel execution.
A generalized output plugin interface can be used to extract any kind of information during the simulation and to visualize the simulation trajectory with MegaMol \cite{Grottel2009, Grottel2010} and other compatible tools.

This paper is organized as follows:
Section \ref{sec:models} describes molecular models which are available in \mardyn.
Section \ref{sec:md-methods} introduces the underlying computational methods.
The implemented load balancing approach is discussed in detail in Section~\ref{sec:parallelization}.
A performance analysis of \mardyn{} is presented in Section~\ref{sec:performance}, including results obtained on two of the fastest HPC systems.

\section{Interaction models in \mardyn}
\label{sec:models}

Molecular motion has two different aspects: External degrees of freedom, corresponding to the translation and rotation with respect to the molecular center of mass, as well as internal degrees of freedom that describe the conformation of the molecule.
In \mardyn, molecules are modeled as rigid rotators, disregarding internal degrees of freedom and employing effective pair potentials for the intermolecular interaction. This modeling approach is suitable for all small molecules which do not exhibit significant conformational transitions. An extension of the code to internal degrees of freedom is the subject of a presently ongoing development, which is not discussed here.
The microcanonical ($NVE$), canonical ($NVT$) and grand-canonical ($\mu{}VT$) ensembles are supported, whereby the temperature is (for $NVT$ and $\mu{}VT$) kept constant by velocity rescaling.

The Lennard-Jones (LJ) potential
\begin{equation}
 U_\text{LJ}(r) = 4\epsilon\left[\left(\frac{\sigma}{r}\right)^{12} - \left(\frac{\sigma}{r}\right)^{6} \right],
\end{equation}
with the size parameter $\sigma$ and the energy parameter $\epsilon$,
is used to account for repulsive and dispersive interactions. It can also be employed
in a truncated and shifted (LJTS) version.\cite{AT87}
LJ potential parameters for the unlike interaction, i.e.\ the pair potential acting between molecules of different species, are determined by the Lorentz and Berthelot combination rules,\cite{Lorentz1881, SVH07, Berthelot1898} which can be further adjusted by binary interaction parameters.\cite{VHH09,SVH03a,VSH05}

Point charges and higher-order point polarities up to second order (i.e.\ dipoles and quadru\-poles), are implemented to model electrostatic interactions in terms of a multipole expansion.\cite{Stone2008, Gray1984}
This allows an efficient computational handling while sufficient accuracy is maintained for the full range of thermophysical properties.\cite{EVH08a}
The Tersoff potential~\cite{Tersoff88} can be used within \mardyn in order to accurately describe a variety of solid materials.\cite{Tersoff89, GVLMFF08}
As a multi-body potential, it is computationally more expensive than electrostatics and the LJ potential.

Any system of units can be used in \mardyn as long as it is algebraically
consistent and includes the Boltzmann constant $\kboltz = 1$ as well as
the Coulomb constant $k_C = 1/(4\pi\epsilon_o) = 1$ among its basic units.
Thereby, expressions for quantities related to temperature and the
electrostatic interactions are simplified. The units of size, energy and charge are
related (by Coulomb's law and the Coulomb constant unit) and cannot be specified
independently of each other. A temperature is converted to energy units by
using $\kboltz$ = $1$, and vice versa. In this way, all other units are determined;
for an example, see \ref{tab:units}.

\begin{table}[h!]
\centering
\caption{A consistent set of atomic units (used by the scenario generators).}
\label{tab:units}
\begin{tabular}{|l|l|} \hline
Boltzmann constant	& $\kboltz = 1$ \\
Coulomb constant	& $k_C = (4\pi\epsilon_0)^{-1} = 1$ \\
& \\
Unit length 		& $l_0 = 1$ $\bohr$ (Bohr's radius) = $5.29177$ $\times$ $10^{-11}$ m \\
Elementary charge	& $q_0 = 1$ $\elementarycharge$ = $9.64854$ $\times$ $10^{9}$ C/mol \\
Unit mass		& $m_0 = 1\,000$ $u = 1$ kg/mol \\
& \\
Unit density		& $\rho_0 = 1 / l_0^{3} = 11\,205.9$ mol/l \\
Unit energy		& $E_0 = k_C q_0^2 / l_0 = 4.35946$ $\times$ $10^{-18}$ J \\
Unit temperature	& $T_0 = E_0 / \kboltz = 315\,775$ K \\
Unit pressure		& $p_0 = \rho_0 E_0 = 2.94211$ $\times$ $10^{13}$ Pa \\
Unit time		& $t_0 = l_0 \sqrt{m_0 / E_0} = 3.26585$ $\times$ $10^{-14}$ s \\
Unit velocity		& $v_0 = l_0 / t_0 = 1620.35$ m/s \\
Unit acceleration	& $a_0 = l_0 / t_0^2 = 4.96148$ $\times$ $10^{-15}$ m/s$^2$ \\
Unit dipole moment	& $D_0 = l_0q_0 = 2.54176$ D \\
Unit quadrupole moment	& $Q_0 = l_0^2q_0 = 1.34505$ D\AA{} \\ \hline
\end{tabular}
\end{table}

\section{Data structures and numerical integration}
\label{sec:md-methods}

The computational core of every MD program is the calculation of forces and torques acting on the molecules, which are based on molecular models for the physical interactions.
The choice of a suitable model depends on many factors, like the material to be simulated, the physical effects studied or the desired accuracy.
Different models may require substantially different algorithms for the numerical solution of Newton's equations of motion.
This can even necessitate major changes in the software structure, e.g.\ when long-range interactions have to be considered explicitly in addition to short-range interactions, or when models with internal degrees of freedom are used instead of rigid molecular models.

In the present version of \mardyn{}, only short-range interactions up to a specified cut-off radius are explicitly computed. The long-range contribution to the pressure and the energy is approximated by isotropic cut-off corrections, i.e. by a mean-field integral for the dispersive interaction, which is supplemented by the reaction field method \cite{Saager1991} for dipolar molecules.



Calculating short range interactions in dynamic systems requires an efficient algorithm for finding neighbors.
For this purpose, \mardyn employs an adaptive linked-cell algorithm.\cite{Buchholz2010}
The basic linked-cell algorithm divides the simulation volume into a grid of equally sized cubic cells, which have an edge length equal to the cut-off radius $\distancecut$.
This ensures that all interaction partners for any given molecule are situated either within the cell of the molecule itself or the $26$~surrounding cells.
Nonetheless, these cells still contain numerous molecules which are beyond the cut-off radius.
The volume covered by $27$ cells is $27$ $\distancecut^3$, whereas the relevant volume containing the interaction partners is a sphere with a radius $\distancecut$, corresponding to $4\pi\distancecut^3\slash{}3$ $\approx$ $4.2$ $\distancecut^3$.
Thus, in case of a homogeneous configuration, only $16 \%$ of all pairs for which the distance is computed are actually considered for intermolecular interactions.

For fluids with computationally inexpensive pair potentials, e.g.\ molecules modeled by a single LJ site, the distance evaluation requires approximately the same computational effort as the force calculation.
Reducing the volume which is examined for interaction partners can therefore significantly reduce the overall runtime.
This can be achieved by using smaller cells with an edge length of e.g.\ $\distancecut/2$, which reduces the considered volume from $27$ $\distancecut^3$ to $15.6$ $\distancecut^3$, so that for a homogeneous configuration, $27 \%$ of the computed distances are smaller than the cut-off radius.

\begin{figure}[ht!]
  \centering
  \includegraphics{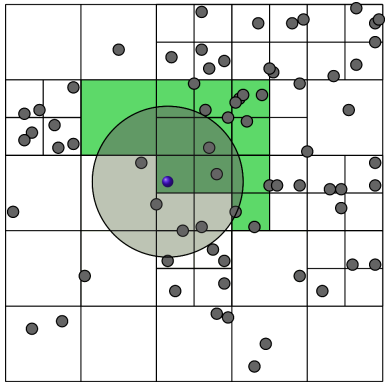}
  \caption{Adaptive cell sizes for an inhomogeneous molecule distribution. Cells that contain significantly more molecules than others are divided into smaller subcells. According to Newton's third law (\textit{actio~=~reactio}), two interacting molecules experience the same force (in opposite directions) due to their mutual interaction, so that a suitable enumeration scheme can be employed to reduce the amount of cell pairs that are taken into account. Following such a scheme, it is sufficient to compute the force exerted by the highlighted molecule on molecules from the highlighted cells.\cite{Buchholz2010}}
  \label{fig:adaptive-cell-sizes}
\end{figure}

However, smaller cells also cause an additional effort, since $125$ instead of $27$ cells have to be traversed.
This is only beneficial for regions with high density, where the cost of cell traversal is small compared to the cost of distance calculation.
Moreover, many applications of molecular dynamics, such as processes at interfaces, are characterized by a heterogeneous distribution of the molecules and thus by a varying density throughout the domain.
To account for this, adaptive cell sizes depending on the local density \cite{Buchholz2010} are (optionally) used in \mardyn, cf.\ \ref{fig:adaptive-cell-sizes}.
Due to periodic boundary conditions, molecules leaving the simulation volume on one side re-enter it on the opposite side, and molecules near the volume boundary interact with those on the opposite side of the volume.

After the calculation of all pairwise interactions, the resulting force and torque acting on each molecule is obtained by summation.
Newton's equations of motion are solved numerically for all molecules to obtain the configuration in the next time step.
Most common methods to integrate these equations are single-step methods, where a new position at the time~\mbox{$t + \delta t$} is calculated from the position, velocity and acceleration at the time~$t$.
This is repeated for a specified number of time steps~$n$ up to the time~\mbox{$t + n \, \delta t$}.
Usually, algorithms based on the (St{\o{}}rmer-)Verlet method \cite{Stoermer1912, Verlet1967} are used.
Instead, \mardyn employs the leapfrog method,\cite{Hockney1970} which is algebraically equivalent to the Verlet method but more accurate numerically.
Positions~${\boldsymbol r}_i$ and velocities~$\dot{\boldsymbol r}_i$ are calculated by
\begin{eqnarray}
\label{eqn:leapfrogv}
\dot{\boldsymbol r}_i \left( t + \frac{\delta t}{2}\right) &=& \dot{\boldsymbol r}_i \left( t - \frac{\delta t}{2}\right) + \delta t \, \ddot{\boldsymbol r}_i(t), \\
\label{eqn:leapfrogr}
\boldsymbol r_i(t + \delta t) &=& \boldsymbol r_i(t) + \delta t \, \dot{\boldsymbol r}_i \left( t + \frac{\delta t}{2}\right).
\end{eqnarray}
For molecules which are not rotationally symmetric, the equations for angular momentum~$\boldsymbol{j}$ and orientation~$\boldsymbol{q}$ (with~$\boldsymbol{q}$ being a quaternion) \cite{AT87} are applied as well.
In analogy to Eqs.~\eqref{eqn:leapfrogv} and \eqref{eqn:leapfrogr} for the translational motion, the rotational motion is described by
\begin{eqnarray}
\label{eqn:leapfrogj}
\boldsymbol j_i \left( t + \frac{\delta t}{2}\right) &=& \boldsymbol j_i \left( t - \frac{\delta t}{2}\right) + \delta t \, \boldsymbol\tau_i(t), \\
\label{eqn:leapfrogq}
\boldsymbol q_i(t + \delta t) &=& \boldsymbol q_i(t) + \delta t \, d\dot{\boldsymbol q}_i\left( t + \frac{\delta t}{2}\right),
\end{eqnarray}
where~$\boldsymbol\tau_i$ is the torque divided by the rotational moment of inertia.

\section{Parallelization and load balancing}
\label{sec:parallelization}

\subsection{Load balancing based on domain decomposition}
A parallelization scheme using spatial domain decomposition divides the simulation volume into a finite number of subvolumes, which are distributed to the available processing units.
Usually, the number of subvolumes and the number of processing units are equal.
This method scales linearly with the number of molecules and is therefore much better suited for large systems than other methods like force or atom decomposition.\cite{Bernreuther2005a,Plimpton1995b, Plimpton1995c}

For heterogeneous scenarios, it is not straightforward that all processes carry a similar share of the total workload.
In simulation scenarios containing coexisting liquid and vapor phases, the local density within the simulation volume can differ significantly, e.g.\ by a factor of $1\,000$.
The number of pairwise force calculations scales quadratically with the density.
Therefore, the computational costs for two subvolumes of equal size may differ by a factor of a million, resulting in many idle processes unless an efficient load balancing scheme is employed.
Depending on the simulation scenario, it may be sufficient to apply a static load balancing scheme which is adjusted only once, or to rebalance the decomposition dynamically, e.g.\ every $100$ to $1\,000$ time steps.

Like in other parts of \mardyn, an interface class is available for the domain decomposition, allowing for the generic implementation of different decomposition approaches and therefore facilitating the implementation of load-balancing strategies based on domain decomposition.
Several strategies were implemented in \mardyn{} and evaluated for nucleation processes.\cite{Buchholz2010}
The strategy based on trees turned out to be the most efficient one. It is available in the current version of \mardyn
as described in the remainder of the present section.

\subsection{Load costs}
\begin{figure}[h]
  \centering
  \subfloat{\includegraphics{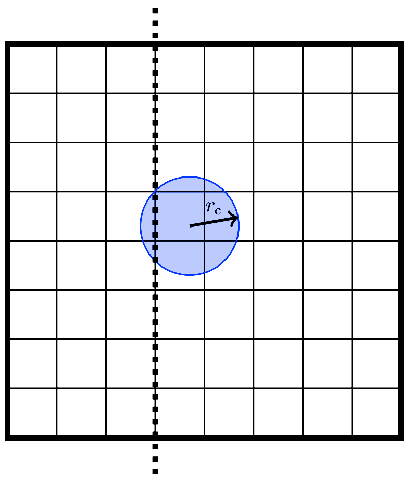}}
  \qquad
  \subfloat{\includegraphics{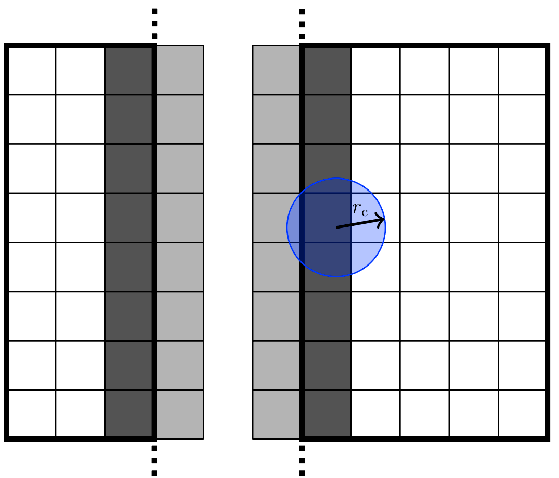}}
  \caption{Left:~The simulation volume (within the bold line) is divided into cells by the linked-cell algorithm (thin lines) where the cell edge length is the cut-off radius $\distancecut$.
  The simulation volume is divided into two subvolumes along cell boundaries (dotted line). Right:~Halo cells (light shaded cells) are introduced storing copied molecule data from adjacent boundary cells (dark shaded cells).}
  \label{fig:decomposition}
\end{figure}

The purpose of load balancing is to decompose and distribute the simulation volume such that all processes need the same computing time.
Such a decomposition requires a method to guess or measure the load that corresponds to a specific subvolume.
The linked-cell algorithm, which is used to identify neighbor molecules, introduces a division of the simulation volume into cells.
These cells are the basic volume units for which the load is determined.
On the basis of the computational cost for each of the cells, a load balancing algorithm can group cells together such that $p$ subvolumes of equal computational cost are created, where $p$ is the number of processing units.
In \ref{fig:decomposition}, a 2D example is given for a simulation volume divided into \mbox{$8 \times 8$}~cells.
This volume is being partitioned along cell boundaries into two subvolumes which will then be assigned to different processes.
The implementation in \mardyn requires each subvolume to cover at least two cells in each spatial dimension.

In a typical simulation, the largest part of the computational cost is caused by the force and distance calculations.
If $N_i$ and $N_j$ denote the number of molecules in cells $i$ and $j$, respectively, the number of distance calculations $n_\mathrm{d}(i)$ for cell $i$ can be estimated by
\begin{equation} \label{eq:load:numdist}
%
  n_\mathrm{d}(i) \approx \frac{N_i}{2} \left(N_i + \sum_{j \,\in\, \mathrm{neigh}(i)} N_j\right).
\end{equation}
The first term in Eq.~\eqref{eq:load:numdist}, i.e.\ $N_i^2/2$, corresponds to the distance calculations within cell $i$. The second term represents the calculation of distances between molecules in cell~$i$ and an adjacent cell~$j$.

While Eq.~\eqref{eq:load:numdist} can be evaluated with little effort, it is far more demanding to predict the number of force calculations. Furthermore, communication and computation costs at the boundary between adjacent subdomains allocated to different process can be significant. They depend on many factors, in particular on the molecule density at the boundary.
Therefore, even if the load on all compute nodes is uniform and remains constant, the location of the subvolume boundaries has an influence on the overall performance.
For a discussion of detailed models for the respective computational costs, the reader is referred to Buchholz.\cite{Buchholz2010} In the present version of \mardyn, the computational costs are estimated on the basis of the number of necessary distance calculations per cell according to Eq.~\eqref{eq:load:numdist}.

\subsection{Tree-based decomposition}
The distribution of cells to processes
is in principle straightforward.
One way is to bring the cells into a linear order (e.g.\ row-wise), walk through the ordered list and sum up the load.
Having reached $1/p$ of the total load, the cells may be grouped together to a subvolume and assigned to a process,
ensuring that all processes carry a similar load.
The problem with this naive approach is that it creates subvolumes with large surface to volume ratios.
A homogeneous system with a cubic volume containing $100 \times 100 \times 100$ cells, distributed to $100$ processes, would for instance be decomposed to $100$ subvolumes with the thickness of a single cell so that all cells would be boundary cells.
In such a case, the additional costs for boundary handling and communication are prohibitively high.

To overcome this problem, a hierarchical decomposition scheme was implemented in \mardyn.
This decomposition is similar to $k$-d trees,\cite{Bentley1975} which are known to achieve a good performance in general simulation tasks \cite{Simon1995} as well as in the special case of particle simulations.\cite{Bernard1999, Fleissner2007}
The simulation volume is recursively bisected into subvolumes with similar load by planes which are perpendicular to alternating coordinate axes.\cite{Berger1987}
To determine the optimal devision plane, the load distribution for every possible division plane is computed and the one resulting in the minimal load imbalance is selected.
This procedure is recursively repeated until a subvolume is assigned to each process.

In case of extremely large simulation volumes, however, initial decompositions are determined following a simplified procedure, until a sufficiently small subvolume size is reached. Thereby, the volume is decomposed into equally sized subvolumes, and the number of processes per subvolume is assigned according to the estimated load for the respective subvolume.

\section{Performance}
\label{sec:performance}

Targeting large-scale parallel runs, \mardyn has been designed for both good single-core and parallel efficiency.
While the code was written in a portable way, which allows to build and execute the program on every standard Linux or Unix system, we focus here on the HPC systems given in \ref{tab:benchmarking-platforms} for the performance analysis. 
In the following sections, we especially explain the influence of the compiler used to build \mardyn on its performance, the overhead of the parallelization as well as its scalability.

\begin{table}[ht]
  \begin{tabular}{|l|l|l|r|}
  \hline
  System, location & Processor type & Interconnect & Cores 
  \\ \hline
  \textit{hermit}, Stuttgart & AMD Opteron 6276                       & Cray Gemini & $113\,664$ \\ 
                             & (Interlagos, 16 cores @\SI{2.3}{GHz})  &             &             \\ \hline
  \textit{laki} (NH), Stuttgart & Intel Xeon X5560                       & InfiniBand  & $5\,600$    \\
                                & (Gainestown, 4 cores @\SI{2.8}{GHz})   &             &              \\ \hline
  \textit{laki} (SB), Stuttgart & Intel Xeon E5-2670                     & InfiniBand  & $3\,072$   \\ 
                                & (Sandy Bridge, 8 cores @\SI{2.6}{GHz}) &             &             \\ \hline
  \textit{SuperMUC}, Garching & Intel Xeon E5-2680                     & InfiniBand  & $147\,456$ \\ 
                              & (Sandy Bridge, 8 cores @\SI{2.7}{GHz}) &             &             \\ \hline
  \end{tabular}
  \caption{HPC platforms used for performance measurements.}
  \label{tab:benchmarking-platforms}
\end{table}

\subsection{Sequential performance}

\begin{figure}[ht]
  \centering
  \subfloat{
     \includegraphics[width=8.33cm]{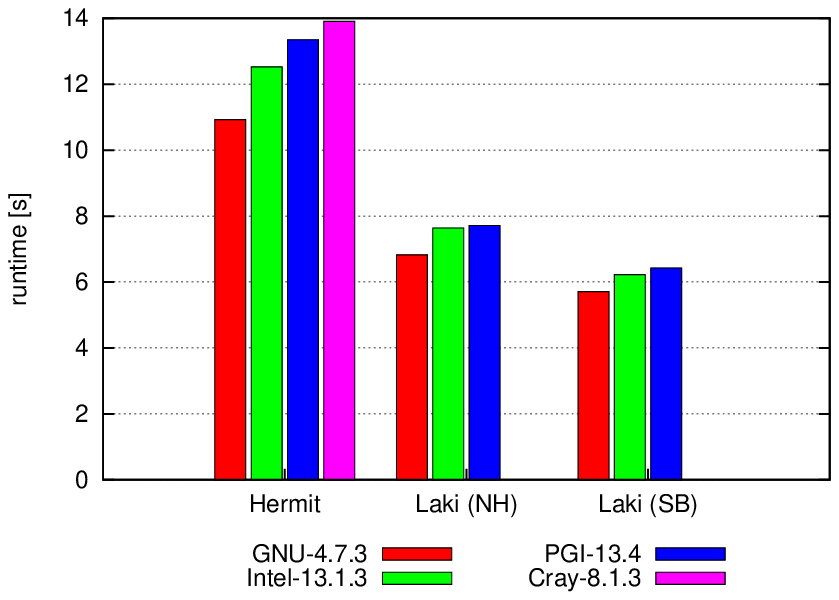}
  }
  ~
  \subfloat{
     \includegraphics[width=8.33cm]{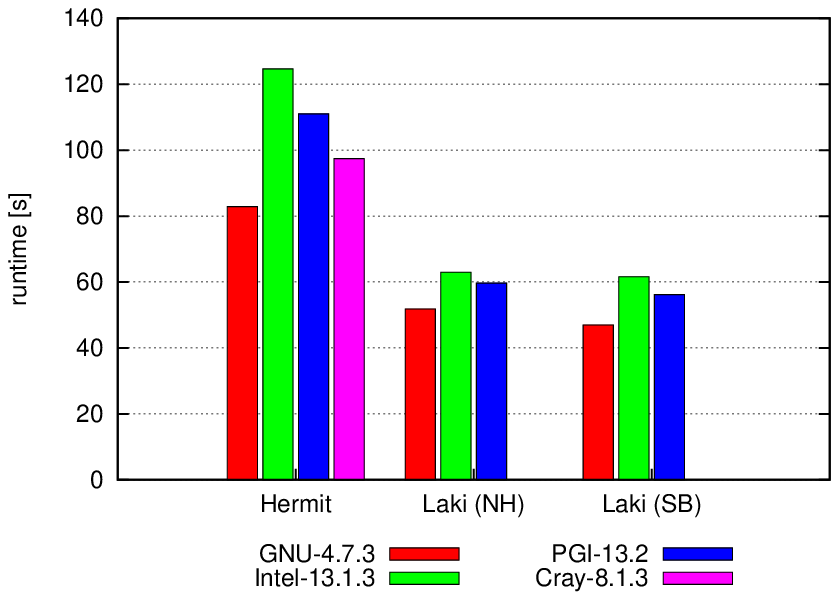}
  }
  \\
  \caption{Sequential execution times of \mardyn on various platforms with different compilers. Scenarios: LJ vapor with $N$ = $40\,000$, $\rho\sigma^3$ = $0.044$, and $kT/\epsilon$ = $0.7$ (left) as well as liquid ethylene oxide with $N$ = $65\,536$, $\rho$ = $19.4$ mol/l, and $T$ = $285$ K (right).}
  \label{fig:compiler-comparison}
\end{figure}

The compiler used to build the code has a large impact on its performance.
\ref{fig:compiler-comparison} shows results obtained with a serial version of \mardyn employing different compilers on the SB and NH partitions of \textit{laki} as well as on \textit{hermit}.
The test scenarios were a LJ vapor (at $kT/\epsilon$ = $0.7$ and $\rho\sigma^3$ = $0.044$) consisting of $40\,000$ molecules and ethylene oxide in a liquid state
(at $T$ = $285$ K and $\rho$ = $19.4$ mol/l) with $65\,536$ molecules.
As can be seen, the sequential program runs fastest on the Sandy Bridge based \textit{laki} system and built with the GNU compiler.
Unless noted otherwise, the GNU~compiler was also used for all further studies discussed below.

\begin{figure}[hbt]
  \centering
  \includegraphics{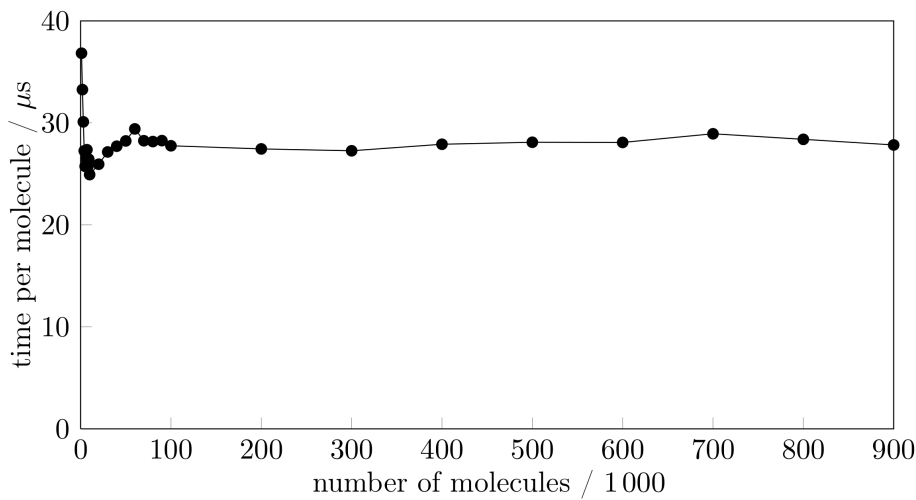}
  \caption{Sequential execution time of \mardyn per molecule, for simulations of a homogeneous LJ fluid at $kT/\epsilon = 0.95, \rho\sigma^3 = 0.6223$ with different system sizes on \textit{laki} (SB).}
  \label{fig:complexity-study}
\end{figure}

The computational complexity of the linked-cell algorithm and domain decomposition scheme used in \mardyn is $\mathcal{O}(N)$.
To evaluate the efficiency of the implementation, runs with different numbers of molecules were performed.
The results in \ref{fig:complexity-study} show that in the present case, the implementation scales almost perfectly with $\mathcal{O}(N)$, as the execution time per molecule is approximately constant.

\subsection{Sequential to parallel overhead}

\begin{figure}[ht]
  \centering
  \subfloat[\textit{Homogeneous liquid} ($N = 2\,048$)]{\includegraphics[width=0.333\textwidth]{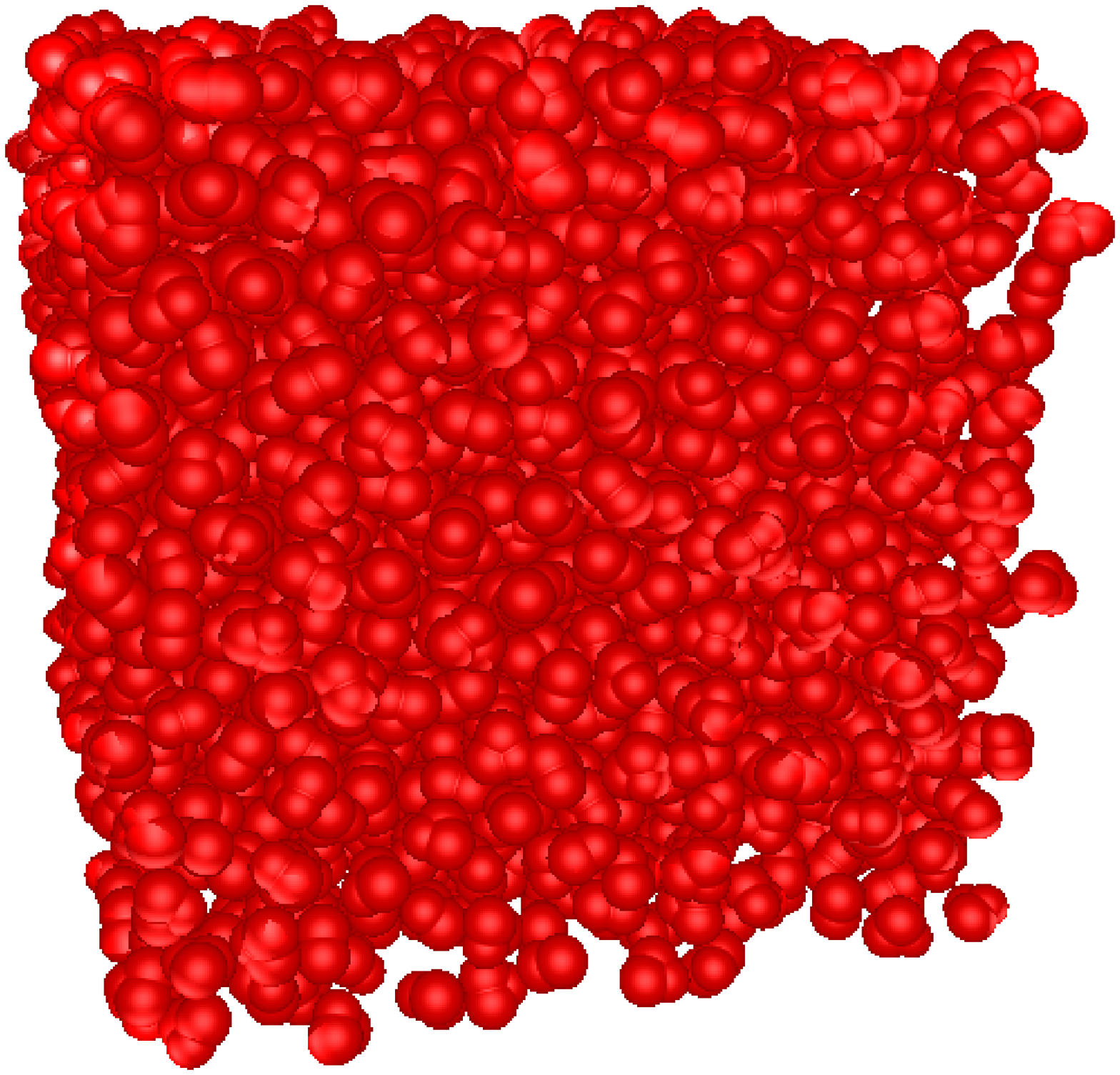}}
  \qquad
  \subfloat[\textit{Droplet} ($N = 46\,585$)]{\includegraphics[width=0.333\textwidth]{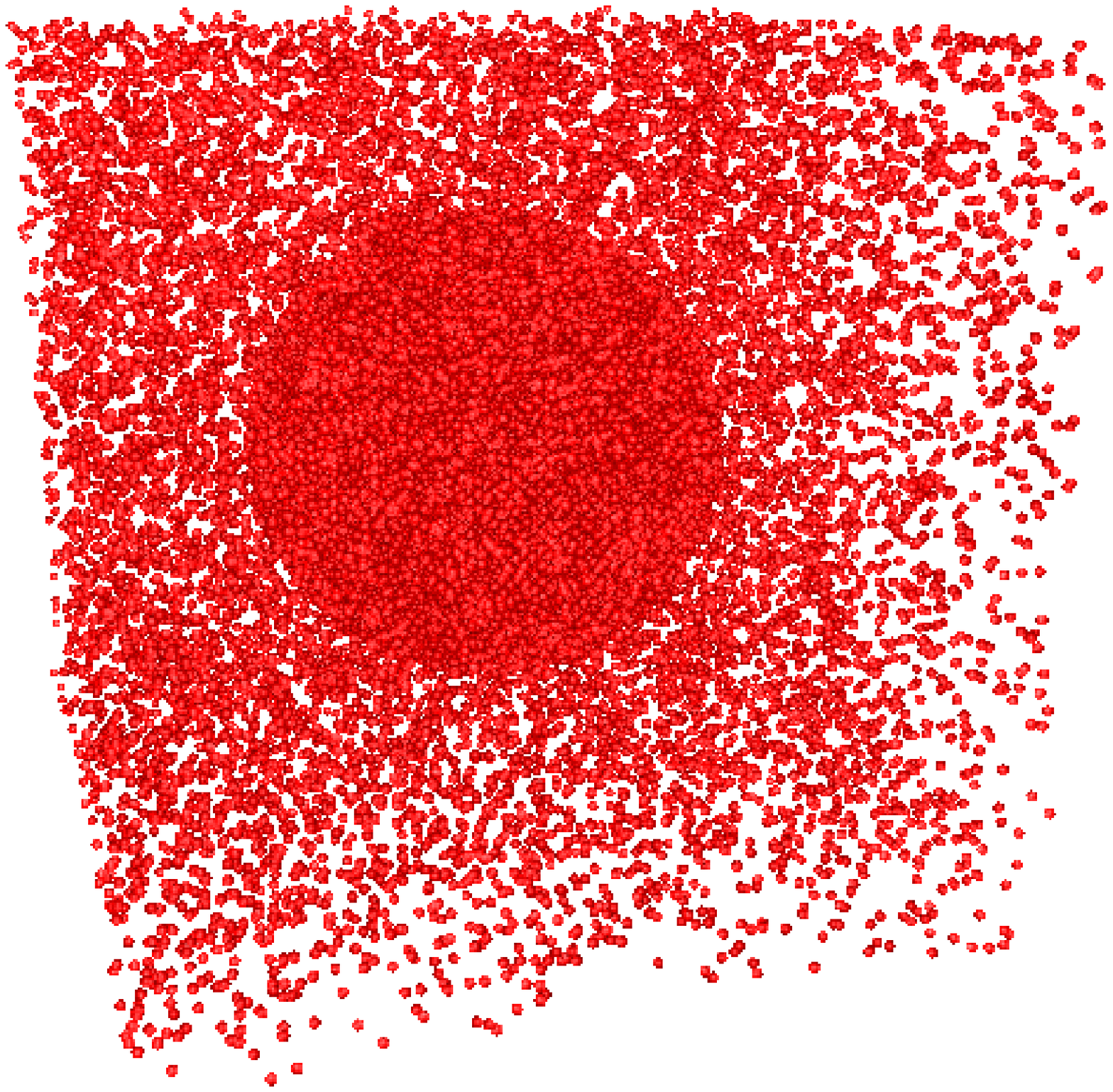}}
  \qquad
  \subfloat[\textit{Planar interface} ($N = 102\,400$)]{\includegraphics[width=0.225\textwidth]{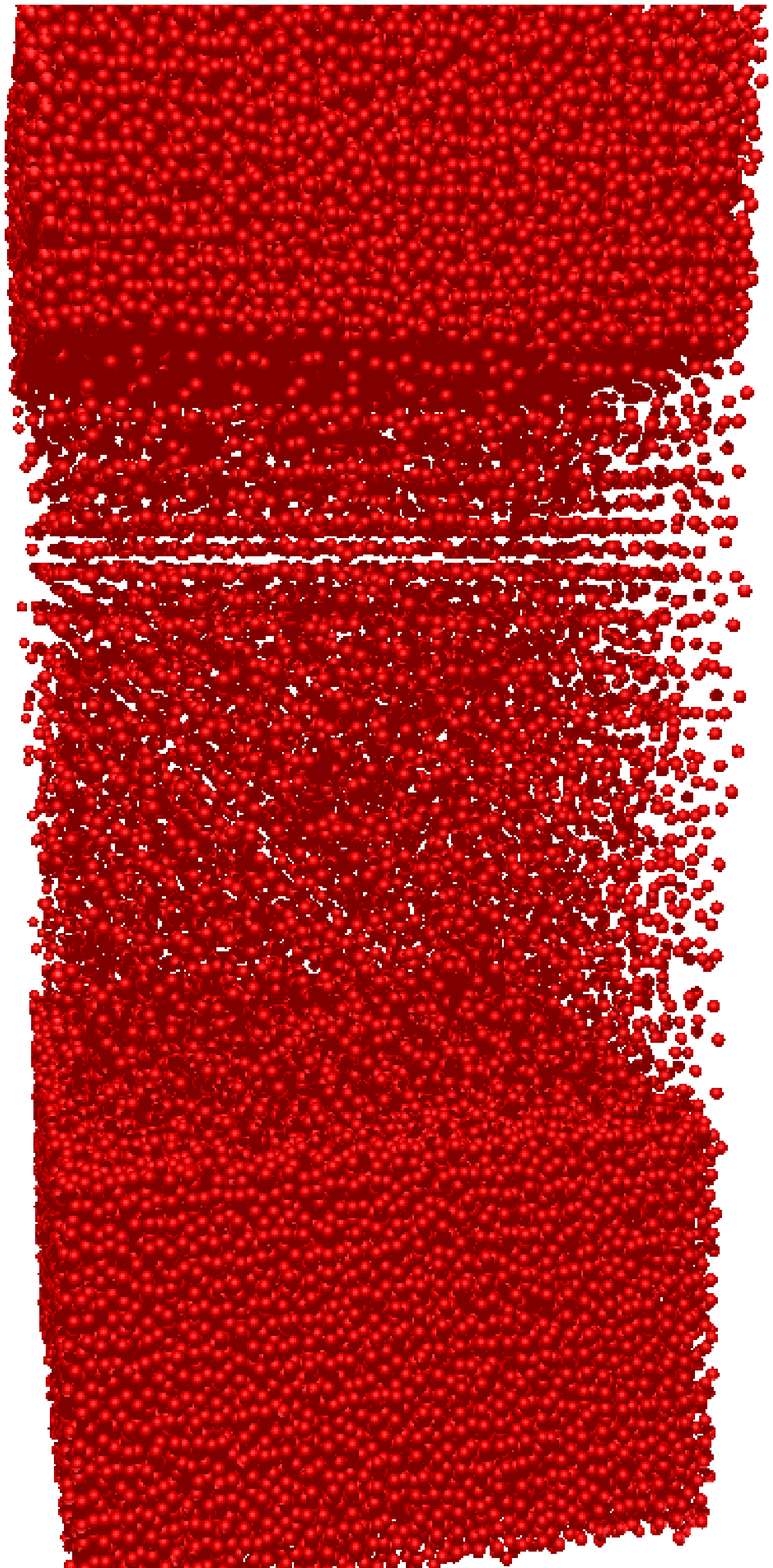}}
  \caption{Scenarios used during the performance evaluation of \mardyn.}
  \label{fig:scenarios}
\end{figure}

For the scalability evaluation of \mardyn, different target scenarios with a varying degree of complexity were considered, produced by the internal scenario generators, cf.\ \ref{fig:scenarios}.
\begin{itemize}
   \item{\textbf{Homogeneous liquid:}} Ethylene oxide at a density of $\rho$ $=$ $16.9$ mol/l and a tempera\-ture of $T = 375$ K. The molecular model for ethylene oxide consists of three LJ sites and one point dipole.\cite{EVH08a}
   \item{\textbf{Droplet:}} Simulation scenario containing a LJTS nanodroplet (cut-off radius $r_\mathrm{c} = 2.5$ $\sigma$) surrounded by a supersaturated vapor at a reduced temperature of \mbox{$kT/\epsilon = 0.95$}.
   \item{\textbf{Planar interface:}} Simulation of a planar vapor-liquid interface of the LJTS fluid (cut-off radius $r_\mathrm{c} = 2.5$ $\sigma$) at a reduced temperature of \mbox{$kT/\epsilon = 0.95$}.
\end{itemize}
In the scenarios, the number of molecules was varied. They were simulated on the platforms given in \ref{tab:benchmarking-platforms} for $1\,000$ time steps and with disabled final I/O.

Parallelization is associated with additional complexity due to communication and synchronization between the different execution paths of the program.
In comparison with sequential execution on a single processing unit, this introduces an overhead.
To determine the magnitude of this overhead for \mardyn, the \textit{planar interface} scenario with $N = 102\,400$ LJ sites was executed over $1\,000$ time steps on the \textit{hermit} system, both with the sequential and the MPI parallel version of the code, but using only a single process.
Execution of the sequential program took $530.9$ s, while the MPI parallel version took $543.4$ s.
This indicates that the overhead due to imperfect concurrency amounts to around $2 \%$ only.

\subsection{Scalability}
\begin{figure}[ht]
  \centering
  \includegraphics{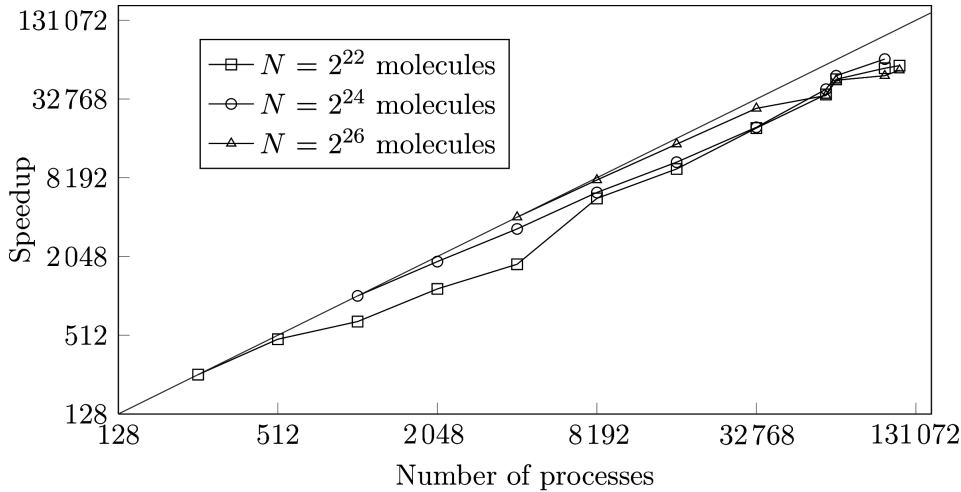}
  \caption{Scaling of \mardyn on \textit{hermit} with the \textit{fluid} example. The starting points of the plots are placed on the diagonal, i.e.\ normalized to a parallel efficiency of 100 \%, neglecting the deviation from perfect scaling for the respective reference case with the smallest number of processes.}
  \label{fig:ls1-scaling-animake-hermit}
\end{figure}

Scaling studies were carried out with the \textit{homogeneous liquid} scenario on the entire \textit{hermit} system, using the standard domain decomposition method, i.e.\ all processes were assigned equal volumes.
The results presented in \ref{fig:ls1-scaling-animake-hermit} show that \mardyn scales favorably in the present case.

\begin{figure}[ht]
  \centering
  {\includegraphics{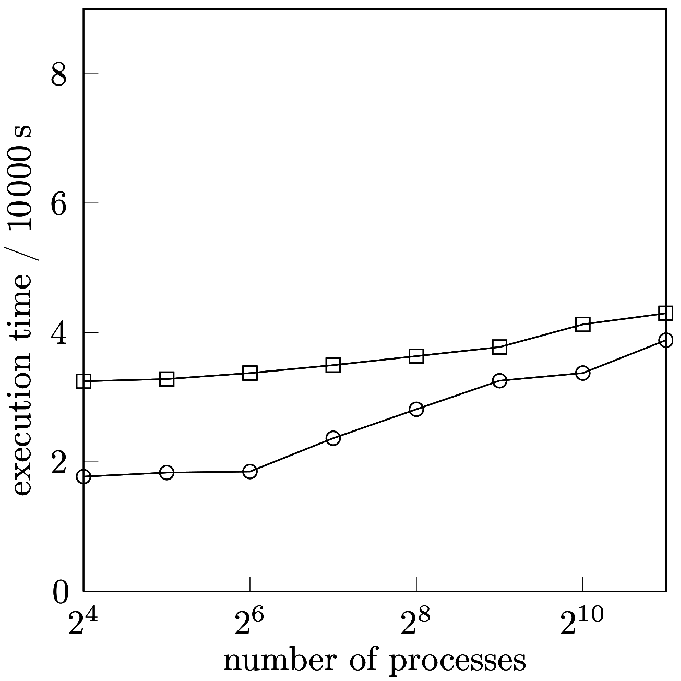}}
  \qquad
  {\includegraphics{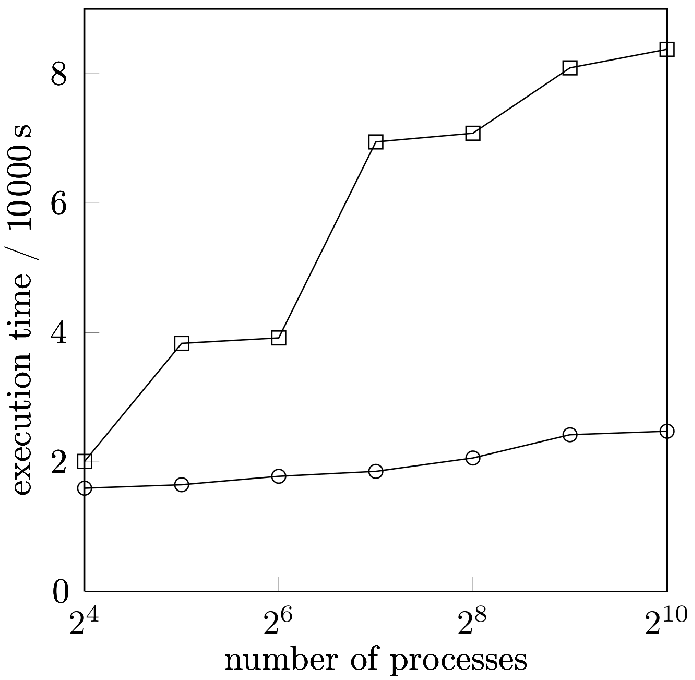}}
  \\
  \caption{Accumulated execution time of \mardyn for a strong scaling experiment on \textit{hermit} using the \textit{planar interface} scenario with $N = 5\,497\,000$ (left) and the \textit{droplet} scenario with $N = 3\,698\,000$ (right). A straightforward static domain decomposition ($\square$), which assigns subdomains with equal volumes to all processing units, is compared with the dynamic $k$-d tree based decomposition ($\circ$).}
  \label{fig:ls1-weak-scaling-inhomogeneous-hermit}
\end{figure}

As discussed above, load balancing is of major importance for inhomogeneous molecule distributions. 
Strong scaling experiments were therefore carried out for the \textit{planar interface} and \textit{droplet} scenarios.
The droplet was positioned slightly off the center of the simulation volume to avoid symmetry effects.
The scenarios were run for $1\,000$ time steps, and the decomposition was updated every $100$ time steps.
The results are presented in \ref{fig:ls1-weak-scaling-inhomogeneous-hermit} and show a clear advantage of the dynamic tree-based decomposition, making the simulation up to four times as fast as the static decomposition into subdomains with equal volume.

\begin{figure}[ht]
  \centering
  \subfloat[Trivial domain decomposition]{\includegraphics[height=0.3\linewidth]{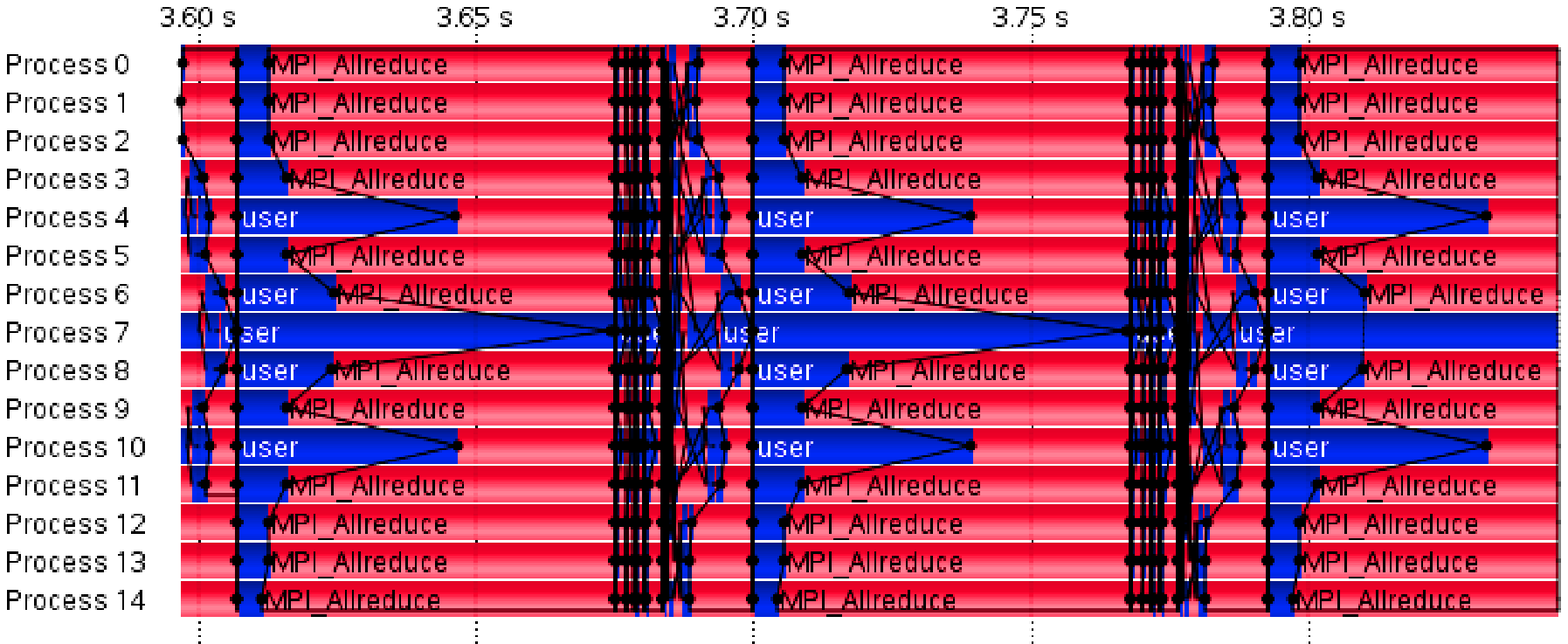}}
  \qquad
  \subfloat[Tree-based domain decomposition]{\includegraphics[height=0.3\linewidth]{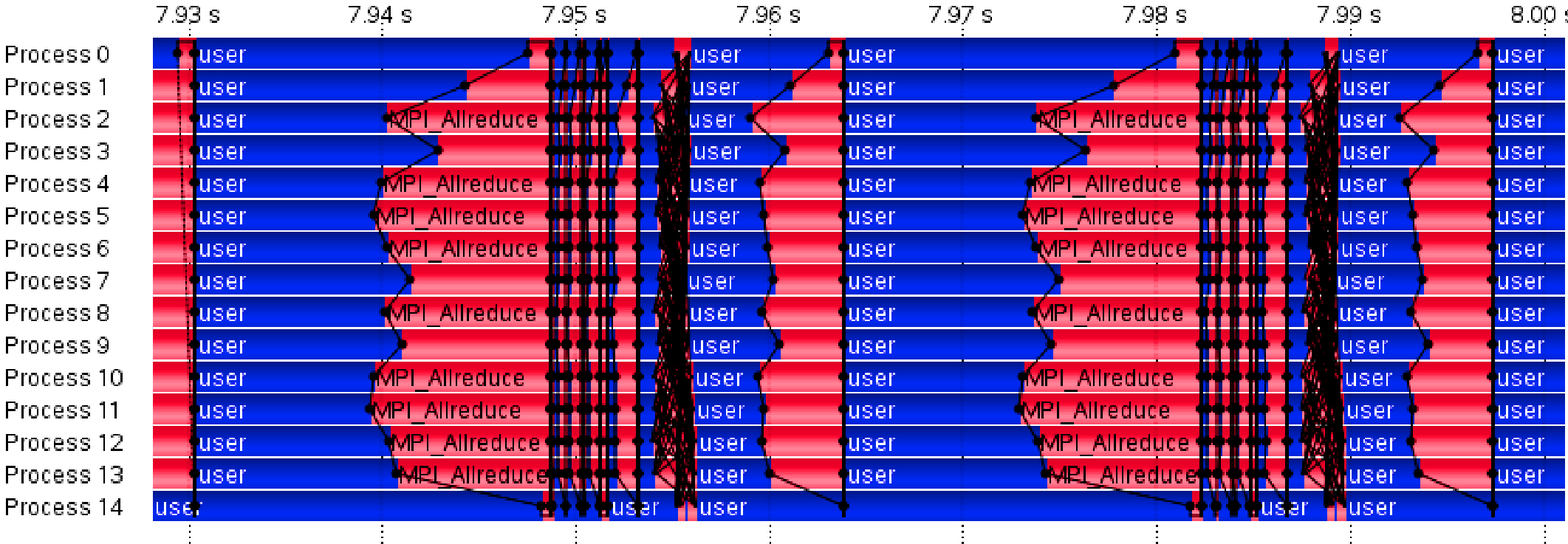}}

  \caption{Traces for the \textit{droplet} scenario on \textit{hermit}, generated with \textit{vampirtrace}. The program state over two time steps is shown for $15$ parallel processes. Computation is indicated by blue colour, communication by red colour. Vertical lines indicate message passing between processes.}
  \label{fig:ls-lb-mkesfera-np14}
\end{figure}

In addition to comparing the run times, the effectiveness of the dynamic load balancing implementation in \mardyn is supported by traces revealing the load distribution between the processes.
\ref{fig:ls-lb-mkesfera-np14} shows such traces, generated with \textit{vampirtrace}, for $15$ processes of a \textit{droplet} scenario simulation on the \textit{hermit} system. For the trivial domain decomposition, $12$ out of $15$ processes are waiting in MPI routines most of the time, while the remaining three processes have to carry the bulk of the actual computation.
In contrast, the $k$-d decomposition exhibits a more balanced distribution of computation and communication.

\subsection{Trillion particle simulation}
\label{sec:triatom}

A version of \mardyn was optimized for simulating
single-site LJ particles on the \textit{SuperMUC} system,\cite{Eckhardt-2013}
one of the largest x86 systems worldwide with $147\,500$ cores and a theoretical
peak performance of more than $3$~PFLOPS.
It is based on a high-performance FDR-10 InfiniBand interconnect by Mellanox and composed of 18 so-called islands,
each of which consists of 512 nodes with 16 Intel Sandy Bridge EP cores at $2.7$ GHz clock speed (turbo mode disabled) sharing $32$ GB of main memory.

Main features of the optimized code version include a lightweight shared-memory parallelization and hand-coded intrinsics in single precision for the LJ interactions within the kernel.
The kernels were implemented in AVX128 (rather than AVX256), mainly for two reasons:
First, the architecture of the Intel Sandy Bridge processor is unbalanced with respect to 
load and store bandwidth, which may result in equal performance for both variants. 
Second, AVX128 code usually shows better performance on the AMD Bulldozer architecture, 
where two processor cores share one 256-bit floating-point unit. 

\begin{figure}[ht]
  \centering
  {\includegraphics{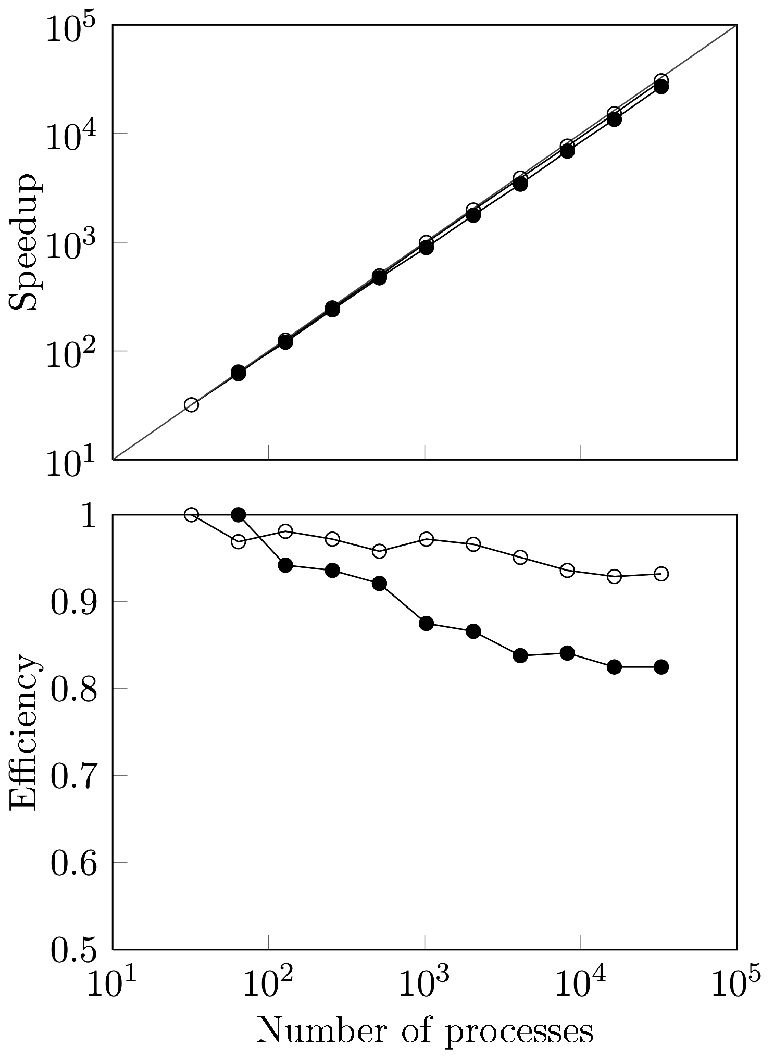}
  }
  ~
  {\includegraphics{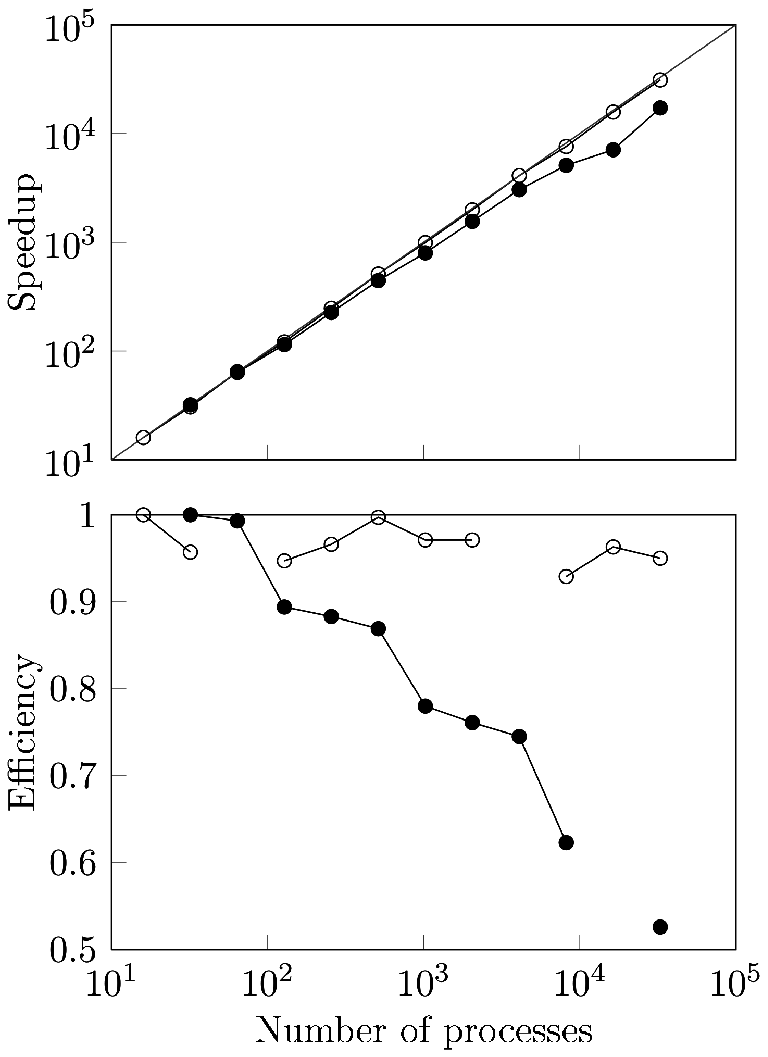}
  }
  \caption{Weak scaling ($\circ$) and strong scaling ($\bullet$) of \mardyn on \textit{hermit} (left) and \textit{SuperMUC} (right), including the speedup (top) and the parallel efficiency (bottom), i.e.\ the speedup reduced by the number of processes. Almost ideal scaling was achieved in case of weak scaling, whereas a parallel efficiency of $53$ \% was obtained in the strong scaling tests on \textit{SuperMUC} and $82.5$ \% on \textit{hermit}, compared to two nodes.}
  \label{fig:scaling}
\end{figure}

To evaluate the performance with respect to strong scaling behavior, a scenario with \mbox{$N$ $=$ $9.5 \times 10^8$}~particles
was studied, which fits into the memory of two nodes, as $18$ GB per node are needed. Thereby, a cut-off radius of $r_c$ $=$ $5$ $\sigma$ was employed.
\ref{fig:scaling} shows that a very good scaling was achieved for up to $32\,768$ cores using $65\,536$ threads. Built with the Intel compiler, the implementation 
delivered a sustained performance of $113$~GFLOPS, corresponding to $8$ \% single-precision peak performance at a parallel efficiency of $53 \%$ compared to $32$~cores ($64$~threads). In addition, a weak scaling analysis with $N=1.6$ $\times$ $10^7$ particles
per node was performed, where a peak performance of $12.9 \%$ or $183$ TFLOPS was achieved at a parallel efficiency of $96$ \% when scaling from $1$ to $32$ $768$ cores.

As the kernel was implemented using AVX128, the same scenario was executed on the Cray XE6 system \textit{hermit} at HLRS, 
however, without shared-memory parallelization and built with the GNU compiler. A noteworthy feature of the Cray XE6 machine is its 3D torus network with Gemini interconnect, which directly plugs in to the HyperTransport 3 host interface for fast MPI communication. On \textit{hermit}, the code achieved a parallel efficiency of $82.5 \%$ and $69.7$ GFLOPS in case 
of strong scaling and $91.5$ \% and $76.8$ TFLOPS or $12.8 \%$ peak performance for weak scaling, respectively,
on $32\,768$ cores in comparison to $64$ cores, i.e.\ two nodes.

As can be seen in \ref{fig:scaling}, the scalability on \textit{hermit} is superior, particularly for strong scaling.
The Gemini interconnect allows for higher bandwidth and lower latency for MPI communications than the FDR-10 InfiniBand
interconnect of \textit{SuperMUC}. Furthermore, a 3D torus network is more
favorable for the communication pattern of \mardyn than the tree topology of \textit{SuperMUC},
where the nodes belonging to each island ($8\,192$ cores) communicate via a fully connected network, while 
for inter-island communication four nodes have to share a single uplink. This can also be seen
in \ref{fig:scaling}, where the scalability noticeably drops when going from 
$8\,192$ to $16\,384$ processes.

As described by Eckhardt et al.,\cite{Eckhardt-2013} a larger weak scaling benchmark on the whole \textit{SuperMUC} was performed with that code version.
Simulating $4.125 \times 10^{12}$ molecules, to our knowledge the largest MD simulation to date,
with a cut-off radius of $r_c$ $=$ $3.5$ $\sigma$, one time step took roughly $\SI{40}{s}$.
For this scenario, a speedup of $133\,183$ compared to a single core with an absolute performance of $591.2$ TFLOPS was achieved, which corresponds to $9.4$~\% peak performance efficiency.

\section{Conclusions}
\label{sec:conclusions}

The massively parallel MD simulation code \mardyn was introduced and presented.
The \mardyn program is designed to simulate homogeneous and heterogeneous fluid systems containing very large numbers of molecules.
Fluid molecules are modeled as rigid rotators consisting of multiple interaction sites, enabling simulations of a wide variety of scenarios from noble gases to complex fluid systems under confinement.
The code, which presently holds the world record for the largest MD simulation, was evaluated on large-scale HPC architectures.
It was found to scale almost perfectly on over $140$ $000$ cores for homogeneous scenarios.
The dynamic load balancing capability of \mardyn was tested with different scenarios, delivering a significantly improved scalability for challenging, highly heterogeneous systems.

It can be concluded that \mardyn, which is made publicly available as free software,\cite{Website} represents the state of the art in MD simulation. It can be recommended for large-scale applications, and particularly for processes at fluid interfaces, where highly heterogeneous and time-dependent particle distributions may occur. Due to the modularity of its code base, future work can adjust \mardyn{} to newly emerging HPC architectures and further extend the range of available molecular modeling approaches and simulation methods. In this way, \mardyn{} aims at driving the progress of molecular simulation in general, paving the way to the micrometer length scale and the microsecond time scale for computational molecular engineering.

\acknowledgement

The authors would like to thank
A.\ Bode and M.\ Brehm for their help in accessing the supercomputing infrastructure at
the Leibniz Supercomputing Center (LRZ) of the Bavarian Academy of Sciences and Humanities.
They thank D.\ Mader for his contribution to developing the very first version of the \mardyn{} program,
S.\ Grottel, M.\ Heinen, D.\ Jenz and G.\ Reina for their work on libraries and tools,
as well as C.\ Avenda\~no Jim\'enez, S.\ Eckelsbach, K.\ Langenbach, R.\ Lustig, S.\ K.\ Miroshnichenko,
E.\ A.\ M\"uller, G.\ Rutkai, F.\ Siperstein, R.\ Srivastava and N.\ Tchipev for fruitful discussions.
The present work was conducted under the auspices of the
Boltzmann-Zuse Society for Computational Molecular Engineering (BZS),
and the molecular simulations were carried out within the supercomputing
project \textit{pr83ri} on the \textit{SuperMUC}
at the LRZ, Garching, and within MMHBF2 on \textit{hermit} and \textit{laki} at the HLRS, Stuttgart.
Financial support is acknowledged due to the IMEMO and SkaSim grants
of the German Federal Ministry of Education and Research (BMBF), and
the Reinhart Koselleck Program as well as the Collaborative Research
Center MICOS (SFB 926) of the German Research Foundation (DFG).


\providecommand{\latin}[1]{#1}
\providecommand*\mcitethebibliography{\thebibliography}
\csname @ifundefined\endcsname{endmcitethebibliography}
  {\let\endmcitethebibliography\endthebibliography}{}


\end{document}